\documentclass[aps,prd,amsmath,amssymb,12pt]{revtex4}
\usepackage{graphicx}% Include figure files
\usepackage{color}
\usepackage{bm}% bold math

\newcommand{\journal}[4]{{#1} {\bf #2}, #3 (#4)}
\newcommand{\be}{\begin{equation}}
\newcommand{\ee}{\end{equation}}
\newcommand{\bea}{\begin{eqnarray}}
\newcommand{\eea}{\end{eqnarray}}
\newcommand{\hf}{\frac12}
\newcommand{\nn}{\nonumber\\}
\newcommand{\eq}[1]{(\ref{#1})}
\newcommand{\la}{\langle}
\newcommand{\ra}{\rangle}

\newcommand{\Tr}{{\mathrm{Tr}}}
\newcommand{\ord}[1]{{\cal O}\left(#1\right)}
\newcommand{\mr}[1]{{\mathrm{#1}}}
\newcommand{\ve}[1]{{\bm{#1}}}
\newcommand{\sign}{\mr{sign}}
\newcommand{\br}{\hskip-6pt/}
\newcommand{\bre}{\hskip-4pt/}

\newcommand{\ddv}[1]{\ddot{\ve{#1}}}
\newcommand{\fd}[2]{\frac{\delta#1}{\delta#2}}

\newcommand{\hx}{{\hat x}}
\newcommand{\hA}{{\hat A}}
\newcommand{\hD}{{\hat D}}
\newcommand{\hG}{{\hat G}}
\newcommand{\ih}{\frac{i}{\hbar}}

\newcommand{\psib}{\bar\psi}

\begin{document}
\title{The Abraham-Lorentz force and electrodynamics at the classical electron radius}
\author{Janos Polonyi}
\affiliation{Strasbourg University, High Energy Theory Group, CNRS-IPHC,23 rue du Loess, BP28 67037 Strasbourg Cedex 2 France}
\date{\today}
\begin{abstract}
The Abraham-Lorentz force is a finite remnant of the UV singular structure of the self interaction of a point charge with its own field. The satisfactory description of such interaction needs a relativistic regulator. This turns out to be a problematic point because the energy of regulated relativistic cutoff theories is unbounded from below. However one can construct point splitting regulators which keep the Abraham-Lorentz force stable. The classical language can be reconciled with QED by pointing out that the effective  quantum theory for the electric charge supports a saddle point producing the classical radiation reaction forces.
\end{abstract}
\maketitle

\section{Introduction}
The radiation reaction problem of point charges in classical electrodynamics, the intrinsic instability of the interaction of a point charge with its own field, has been clearly stated since more than a century, however the discovery of quantum mechanics somehow deflected the interest of the majority of the physics community. Nevertheless number of ideas and methods have been presented to tackle the problem and several solutions have been proposed in the meantime. The present work is based on a new point of view, namely the proper derivation of a causal and stable radiation reaction force of point particles needs an UV cutoff even in classical field theory. 

Let us approach the radiation reaction problem, the last open chapter of classical electrodynamics, in three steps:

1. {\em Origin}: It is known that an accelerated charge looses part of its energy by radiation hence there is a radiation reaction force. However the Lagrangian does not contain higher than first order time derivatives hence can not give account the energy, lost to the radiation. The Abraham-Lorentz force can nevertheless be identified by checking the energy-momentum balance equation for the radiating charge \cite{rohrlich65,jackson}. 

2. {\em Unacceptable features}: The Abraham-Lorentz force is an $\ord{\dddot x}$ dissipative force which generates self-accelerating, runaway trajectories. Such a state of affairs questions the applicability of field theory to point particles and a wide range of remedies have been proposed. (i) The problem is related to the third auxiliary condition, needed to integrate the equation of motion. By imposing an additional final condition one trades the instability into acausality \cite{dirac}, an idea which can be realized by assuming a complete absorption of the electromagnetic radiation at the final state \cite{wheeler}. (ii) One can expand the solution in the retardation and the Abraham-Lorentz force can be approximated by making an iterative step, using the second order equations. The result is an $\ord{\ddot x}$ equation which remains stable as long as the resolution is worse than $r_0$ \cite{landau,rohrlichll}. The same equation can be recovered by restricting the trajectories of the Abraham-Lorentz force to the stable manifold \cite{spohn}. (iii) A more flexible family of modifications results from giving up the local nature of the reaction force in time. One way to achieve this is to retain a memory term in the reaction force and the resulting integro-differential equation offers important improvements \cite{nonloceq}. Another possibility is to assume and extended charge distribution \cite{al} of non-electromagnetic origin \cite{extch,levine} or the presence of a polarizable medium \cite{blinder}. Yet another approach is to assume a suitable chosen form factor \cite{formf}. A particular non-locality has been evoked by assuming a discrete structure in time \cite{caldiriola}. (iv) Finally, one can step back and seek a change of the effective equation of motion, based on physical intuition \cite{change}, on magnetic moment charge \cite{magneticch}, or on quantum effects \cite{quant}. So far no generally accepted solution has been found to cure the instability. 

3. {\em Semiclassical effect}: The radiation reaction problem seems to be a problem of classical electrodynamics. However it appears at a length scale, the classical electron radius which is close to the size of a proton. Thus the proper derivation of the real radiation reaction force has to explain the emergence of a classical phenomenon deeply within the quantum domain. The Abraham-Lorentz force can be identified in the $\hbar\to0$ limit of the shift of the position of a wave packet \cite{higuchi} and the Landau-Lifsitz form of the equation \cite{landau} follows from considering the shift of the momentum of the particle \cite{krivitskii}. The more direct derivation which goes parallel with the classical arguments is based on the retarded Green's function. It has already been considered in the presence of a cutoff \cite{coleman} but the non-locality prevented a systematic, analytical exploitation of the equation of motion. 

The strategy, followed in this work, is borrowed from the renormalization group method and is based on the use of the effective action. The solution of the problem, mentioned in point 1. above, is the realization that the radiation reaction arises as an effective force. In fact, the construction of the effective charge dynamics by solving Maxwell's equations for a given particle world line and inserting the solution back into the mechanical equation of motion generates the Abraham-Lorentz force \cite{point}. Our procedure is minimalist in the sense that the problems about the Abraham-Lorentz force are solved in electrodynamics, without invoking other interactions and is close to the family of propositions (iii), mentioned in point 2. above. The emergence of a semiclassical phenomenon at the classical electron radius, the issue of point 3., is justified by pointing out that the radiation reaction force is an $\ord{\hbar^0}$ saddle point effect of QED. The construction of the effective charge dynamics includes several novel features:

A. {\em Loop integrals in classical theories}: We are accustomed to UV finite classical theories because the iterative solution of the classical equations of motion can be presented in the form of series, involving $\ord{\hbar^0}$, tree-level graphs. But this holds for elementary, closed theories only. If there are unobserved degrees of freedom, an environment, and we are looking for a closed system of equations for the observed quantities, then the elimination of the unobserved coordinates generates loop diagrams. The iterative solution of the classical equation of motion of a charge, moving on the background of a fixed electromagnetic field, can be represented by an infinite series of tree-level graphs where the electromagnetic field is attached to the world line as an external leg. However the elimination of the electromagnetic field by the help of the Maxwell equations, using the world line as a source, couples the pairs of the external legs and forms loops. The loop integral is $\ord{\hbar^0}$, a power of $\hbar$ is lost compared to the usual counting in QED because the line, representing the charge is classical, $\ord{\hbar^0}$. The loop integrals of classical effective theories have already been spotted as the on-shell contributions to the loop integrals of the full quantum theory \cite{holstein}. 

B. {\em Regularization of classical theories:} The self-interaction of a point charge with its own field is UV singular due to the $\ord{r^{-1}}$ near field, requiring the usual regularization and renormalization procedure of quantum filed theories, applied even in classical field theories. The appearance of the cutoff in the classical effective equation of motion makes certain concepts of the renormalization group method important for classical field theories.

C. {\em Non-local saddle point in quantum theories:} The emergence of an $\ord{\hbar^0}$ loop-integral in quantum field theories indicates the presence of a saddle point \cite{qrg}. This saddle point is inhomogeneous and represents a non-local interaction. This can be understood by recalling that on the one hand locality stands for the analycity in the momentum variables at $p^\mu=0$ and the other hand the saddle point, the on-shell radiation field, is proportional to a Dirac-delta hence is non-analytic in the momentum space.

D. {\em Modified conservation laws in non-local theories}: The cutoff renders the dynamics non-local at the cutoff scale. The Noether currents are not conserved in non-local theories however they can be made to satisfy the continuity equation by adding appropriate non-local terms \cite{stab}. These terms are not definite and render the conserved energy unbounded from below, casting doubt on the stability of the cutoff theory. The traditional strategy to derive the effective equation of motion is based on the conservation of the energy-momentum of the unregulated, naive theory \cite{dirac}. Such an argument fails even if it is based on the energy-momentum flux, calculated at finite distance from the charge, on the surface of a tube around the point particle world line.

E. {\em Open effective theory}: The dynamics of the regulated cutoff theory is open owing to the unobserved UV degrees of freedom hence it should be recast in a framework designed for open systems \cite{qrg}, namely within the Closed Time Path (CTP) formalism \cite{schw}. This is a CQCO scheme, it handles classical, quantum, closed and open systems on equal footing and treats initial rather than boundary value problems. The redoubling of the degrees of freedom, the distinguishing feature of this scheme, allows the extension of the variational principle of classical mechanics for dissipative forces in open systems \cite{cctp} and the quantum effects arise as an $\cal O(\sqrt\hbar)$ separation of the two coordinates, describing the same degree of freedom. Furthermore we obviously have to rely on initial rather than boundary conditions in problems, related to the radiation. 

There is yet another reason to use the CTP scheme: The saddle point mentioned in point C. is suppressed in the usual formalism of quantum field theory by imposing the vacuum as the final state. To retain the emitted radiation one has to allow arbitrarily many on-shell excitations in the final state which can be achieved only within the CTP formalism. 

F. {\em Stable relativistic cutoff}: The radiation reaction is a semiclassical effect and the $\ord{\hbar^0}$ effective action needs a regulator. This regulator has to be Lorentz invariant \cite {lorentzsens} because the diverging near field at the point charge amplifies the eventual symmetry breaking effects. The choice of a relativistic regulator with stable dynamics is rather difficult due the inherent relation between boost symmetry and divergent loop integrals \cite{psplithd}, resulting from the infinite volume of the boosts within the Lorentz group. However the free boost invariant cutoff theories are stable because the unstable channels are opened by the interactions only. Hence the stability can be achieved by an appropriate suppression of the interactions, provided by relativistically invariant point-splitting. The regulator, used in this work, consists of the smearing of the electromagnetic field in the interaction term. The smearing is chosen in such a manner that charges, moving with the speed of light do not interact. The resulting effective dynamics is stable since a runaway charge must acquire velocities beyond the speed of light. 

The organization of this paper is the following: The classical calculation is presented in section \ref{cleffths}. First we derive the linearized effective equation of motion of classical, regulated electrodynamics \cite{point} in section \ref{lineoms}, describing a stable dynamics for low cutoff. If the resolution in the space-time is better than the classical electron radius then the usual $\ord{\dddot x}$ Abraham-Lorentz force induces self accelerating, unstable particle trajectories. Next we turn to the full, non-linear, integro-differential equation of motion in section \ref{fulleffs}. Its numerical solution displays two distinct scale regimes, separated by the classical electron radius. The usual linear Abraham-Lorentz force is displayed in the IR domain and the non-linear effects prevent the run away and keep the motion stable in the UV regime. In the second part of the paper, section \ref{saddlepds}, the issue of embedding such a classical scenario in QED is taken up briefly. The scale hierarchy of QED is commented in section \ref{scaless} to argue in favor of an island of semiclassical saddle point physics around the classical electron radius scale within the quantum domain, followed by the derivation of the effective dynamics for the electric current in section \ref{effactpds}. The point splitting regularization is outlined in section \ref{regs}. Finally, the effective action, leading to the classical equations of motion of section \ref{cleffths} is presented in section \ref{wlacts}. The conclusions are summarized in section \ref{concls}.

\section{Classical effective dynamics of a point charge}\label{cleffths}
The discussion of the saddle point of the emitted radiation can be carried out within classical electrodynamics and the derivation of the regulated classical action from QED is postponed until section \ref{effactpds}. All we need to know at this point to embark the classical calculation is the regulator: It  consists of the replacement of the Dirac-delta in the retarded Green's function of the electromagnetic field, $D^r_{BCl\mu\nu}(x)$ where the indices $B$ and $Cl$ indicate that this is the bare, regulated Green's function with the classical electromagnetic $4\pi$ convention, by another function to be specified below, $\delta(x^2)\to\delta_B(x^2)$. 

The effective dynamics of a point charge is found by solving Maxwell's equation for the electromagnetic field, $A_\mu(x)$, induced by a point charge which follows a given world line, $x^\mu(s)$, and substituting the result into the equation of motion of the charge. The electromagnetic field,
\be\label{indgf}
A_\mu(x)=e\int dsD^r_{BCl\mu\nu}(x-x(s))\dot x^\nu(s),
\ee
is given by the transverse retarded Green's function, $D^r_{BCl\mu\nu}(x)=-4\pi(g_{\mu\nu}-\partial_\mu\partial_\nu/\Box)D^r_0(x)$, where $D^r_0(x)=-\Theta(x^0)\delta_B(x^2)/2\pi$ is the free, massless Green's function and $\dot x(s)=dx(s)/ds$. The mechanical equation of motion,
\be\label{meom}
m_Bc\ddot x^\mu=\frac{e}c[\partial^\mu A^\nu(x)-\partial^\nu A^\mu(x)]\dot x_\nu-k^\mu,
\ee
contains the bare mass, $m_B$, and an external source, $k^\mu(s)$, to diagnose the dynamics. 

The regulated Dirac-delta should satisfy three conditions: (i) To preserve the flux of the radiated field we require the normalization condition
\be\label{regk}
\int_{-\infty}^\infty dz\delta_B(z)=1.
\ee
(ii) The Lorentz symmetry prevents us to smear the factor $\Theta(x^0)$ of the retarded Green's function,  we impose $\delta_B(0)=0$ to separate the singular points in the product of two generalized functions in the Green's function. (iii) The stability of the dynamics requires the suppression of the interactions for charges moving faster than the light, expressed by the condition $\delta_B(z)=0$ for $z<0$. The simplest possibility is to displace its retarded Green's function slightly off the light-cone,
\be\label{shdd}
\delta_B(x^2)=\delta(x^2-\ell^2),
\ee
It leads to oscillations in the momentum space which can be avoided by smearing the singularity, 
\be\label{gdd}
\delta_B(x^2)=\frac{\Theta(x^2)}{12\ell^4}x^2e^{-\frac{\sqrt{x^2}}\ell}.
\ee

The effective equation of motion with the regulated self interaction is 
\be\label{fulleom}
\ddot x=4r_{0B}\int_{-\infty}^sds'\delta'_B((x-x')^2)\{(x-x')(\dot x\dot x')-[\dot x(x-x')]\dot x']\},
\ee
where $x=x(s)$ and $x'=x(s')$, and $r_{0B}=e^2/m_Bc^2$ stands for the bare classical electron radius and $\delta'(z)=d\delta(z)/dz$. The characteristic scale of the radiation reaction problem is provided by the only dimensional constant of this equation, the classical electron radius, playing the role of the coupling constant.

\subsection{Linearized equation of motion}\label{lineoms}
The linearized the equation of motion,
\be
\ddot x=4r_{0B}\int_{-\infty}^0du\delta'(u^2)(x-x'+u\dot x'),
\ee
contains a regular and a singular force on the right hand side,
\bea
F_r&=&2r_{0B}\int_{-\infty}^0\frac{du}{u^2}\delta(u^2)\left[x-x'+u\dot x'-u^2\left(\ddot x'-\hf\ddot x\right)+\frac{2u^3}3\dddot x\right],\nn
F_s&=&-r_{0B}\int_{-\infty}^0du\delta(u^2)\left(\ddot x+\frac{4u}3\dddot x\right)
\eea
where the expression in the square bracket of the integrand of $F_r$ contains the regular, $\ord{u^4}$ terms making up a uniformly convergent integral. The rest, $F_s$, can be written as
\be
F_s=-\frac{r_{0B}}2\ddot x\int_0^\infty\frac{dz}{\sqrt{z}}\delta_B(z)+\frac23r_{0B}\dddot x,
\ee
giving rise to a mass renormalization, $m=m_B+\delta m$, with
\be\label{massct}
\delta m=\frac{e^2}{2c^2}\int_0^\infty\frac{dz}{\sqrt{z}}\delta_B(z).
\ee
and to the Abraham-Lorentz force. 

It is illuminating to check the order of magnitude of the two forces as the cutoff is removed. We assume that $\Lambda=1/\ell$ is large enough to approximate the square bracket in the integrand of the uniformly convergent part by $u^2/\ell_x^3$ where $\ell_x$ is the length scale of the world line and find
\be
F_r\approx2\frac{r_{0B}\ell}{\ell^3_x}\int d\tilde u\tilde\delta_B(\tilde u^2)\tilde u^2,
\ee
in terms of the dimensionless variable $\tilde u=u/\ell$ and the regular function $\tilde\delta_B(\tilde u^2)=\ell^2\delta_B(u^2)$. The regular force is suppressed during the renormalization owing to the smallness of the important integration region, $u=\ord\ell$ of an $\delta(u^2)u^2=\ord{\ell^0}$ integrand. The important integration region is $u=\ord\ell$ in the non-uniformly convergent part, too. The linearly divergent mass renormalization comes from an $\ord{\ell^{-2}}$ integrand and the reduced $\ord{\ell^{-1}}$ divergence of the integrand in the Abraham-Lorentz force yields a cutoff independent result. Such a cutoff-independent cutoff-scale contribution owes its existence to the non-uniform convergence and is the hallmark of anomalies. In fact, the coefficient of $\dddot x$ is an ``accidentally finite'' loop-integral as in the case of the chiral anomaly for massless fermions.

The retarded world line Green's function, $F^r$, is defined by
\be
\dot x^\mu(s)=\int_{-\infty}^\infty ds'F^r(s-s')k^\mu(s'),
\ee
and its Fourier transform,
\be
F^r_\omega=\int_{-\infty}^\infty dse^{i\omega s}F^r(s),
\ee
can be written as $F^r_\omega=1/[(\omega+i\epsilon)^2\chi^r_\omega]$, in terms of the susceptibility 
\bea\label{worldg}
\chi^r_\omega&=&1+r_0\biggl[i\frac23\omega\\
&&-\frac2{\omega^2}\int_{-\infty}^0du\delta_B(u^2)\frac{(1+i\omega u-u^2\omega^2)e^{-iu\omega}-1+\hf u^2\omega^2-i\frac23u^3\omega^3}{u^2}\biggr],\nonumber
\eea
and the long memory tail when the regulator \eq{gdd} is used requires $\mr{Im}\omega>0$. The harmonic effective dynamics is stable and causal if the susceptibility is analytic and pole-free on the upper part of the complex $\omega$ plane. 

The rational function, multiplying the Dirac-delta in the integrand  is $\frac38\omega^4u^2(1+\ord{\omega u})$ hence
\be\label{dimlesseig}
\chi^r_\omega=1+r_0\omega\left[\frac23i+\ord{\omega\ell}\right].
\ee
As the cutoff is removed with a fixed frequency the Abraham-Lorentz force is left behind and the infamous self-acceleration is recovered. The unstable dynamics generates high frequency components which make the particular details of the regulator influence the motion. The numerical monitoring of the $\ell$-dependence shows that both regulators lead to a susceptibility which develops a pole with positive imaginary part, destroying the stability when the cutoff is comparable or larger than the classical electron radius, as predicted by the cutoff-independent acausal pole of the susceptibility \eq{dimlesseig}.

One would think that the appearance of the instability at $\ell\sim r_0$ implies that the source of the instability is related to the dynamics around the scale $r_0$. But the dynamics is shaped both by the $\ord{\omega^2}$ kinetic energy and the self interaction. The $\ord\omega$ friction term is forbidden by Lorentz symmetry, the impossibility of measuring absolute velocity, thus the radiation energy loss is represented by the $\ord{\omega^3}$ dissipative force. This latter which is weak for slow motions turns out however to be dominant at high frequencies, the crossover being around $\omega\sim1/\ell$ and makes the linearized dynamics unstable in the UV regime.

\subsection{Full effective dynamics}\label{fulleffs}
The regulator was introduced    in such a manner that the interaction becomes suppressed if the velocity of the particle exceed the speed of light, making the instability, arising from the self interaction, unable to drive the particle to arbitrarily high velocity. Therefore it is natural to inquire whether the non-linear terms of the effective equation of motion \eq{fulleom} can stabilize the dynamics. The traditional derivation of the effective equation of motion, based on the energy-momentum conservation, leads to an equation where the only non-linearity arises from the projection of the reaction force onto the linear subspace, orthogonal to the four-velocity \cite{rohrlichajp0}. The cutoff-dependence of the instability suggests that if the non-linearities stabilize the dynamics they should come from another source than this cutoff-independent projection operator. 

The shifted Dirac-delta, \eq{shdd}, produces the equation of motion with a finite delay,
\bea\label{fullsh}
\ddot x&=&r_0\frac{m}{m_B}\frac1{[\dot x'(x-x')]^2}\biggl[\frac{\ddot x'(x'-x)+1}{\dot x'(x-x')}\{(x-x')(\dot x\dot x')-[\dot x(x-x')]\dot x'\}\nn
&&+(x-x')(\dot x\ddot x')+[\dot x(x'-x)]\ddot x'\biggr],
\eea
where the retarded source point, $x'$, is found by the condition $\ell^2=(x-x')^2$. The equation of motion, found by the help of the smeared Dirac-delta, \eq{gdd}, has infinitely long memory and contains the velocities only in the right hand side,
\bea\label{fullsm}
\ddot x&=&\frac{r_0}{3\ell^4}\frac{m}{m_B}\int_{-\infty}^0du\left(1-\frac{\sqrt{(x-x')^2}}{2\ell}\right)e^{-\frac{\sqrt{(x-x')^2}}\ell}\nn
&&\times\{(x-x')(\dot x\dot x')+[\dot x(x'-x)]\dot x']\}.
\eea
We impose the initial condition that the charge is at rest, $x(t)=(t,\ve{0})$, for $t<t_0$ and the charge follows a prescribed trajectory, $\ve{x}_i(t)$ for $t_0<t<t_0+t_i$. A certain external source, $k_i(s)$, is supposed to generate this motion which is turned off after this initial phase and the invariant length, $s$, of the world line is measured from the time $t_0+t_i$. The numerical solution of the equation of motion becomes straightforward with such initial conditions: One introduces a small but finite $\Delta t$ step size and writes eqs. \eq{fullsh} or \eq{fullsm} as differential equation and finds the retarded time, $u$, or calculates the integral of the memory term numerically, respectively at each step, $t\to t+\Delta t$.

\begin{figure}
\includegraphics[scale=.7]{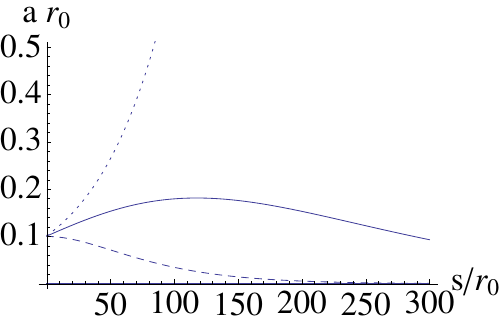}\hskip.5cm\includegraphics[scale=.7]{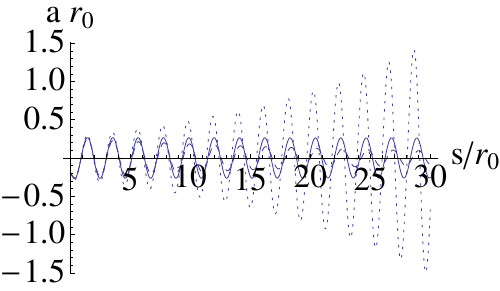}

\centerline{(a)\hskip7cm(b)}
\caption{A component of the spatial acceleration, $a r_0=|\ddv{x}|r_0$, plotted against the proper time, $s/r_0$ for the smeared Dirac-delta regularization, $r_0/\ell=3$, (a): $m/m_B=1.95$ (dashed line), $m/m_B=1.98$ (solid line) and $m/m_B=2$ (dotted line) and (b): $m/m_B=-3.8$ (dashed line), $m/m_B=-3.91$ (solid line) and $m/m_B=-4.1$ (dotted line).}\label{stabilf}
\end{figure}

The equation of motion has two free, adjustable parameters, the cutoff, $\ell$, and the bare mass, $m_B$. Hence we need a renormalization condition to fix the theory, it is chosen to be 
\be\label{rencond}
\chi^r_\omega=1+\frac23ir_0\omega,
\ee
cf. eq. \eq{dimlesseig}.

The numerical solution of the equation of motion indicates stable dynamics for sufficiently weak force, i.e. small $|m/m_B|$. The acceleration changes in a monotonous, exponential manner after some transient period, depending on the initial conditions, if $m_B>0$ as shown in Fig. \ref{stabilf} (a) and the renormalization condition, \eq{rencond}, can be satisfied by monitoring the relaxation for large $s$. When $m_B<0$ then the acceleration is oscillatory with exponentially exploding or decreasing envelope, cf. Fig. \ref{stabilf} (b). The relaxation of the envelope is used to find the physical theory, obeying eq. \eq{rencond} in this case. The precise value of $m/m_B$ at the stability edge is found to be slightly dependent on the initial, prescribed trajectory. This might come from the finite $\Delta s$ resolution of the finite difference equation, solved numerically because the unstable, runaway trajectories support no fixed, finite $\Delta s$. The existence of stable regions suggests that despite the unboundedness of the energy in regulated electrodynamics there are energy barriers which stabilize the charge.

\begin{figure}
\includegraphics[scale=.7]{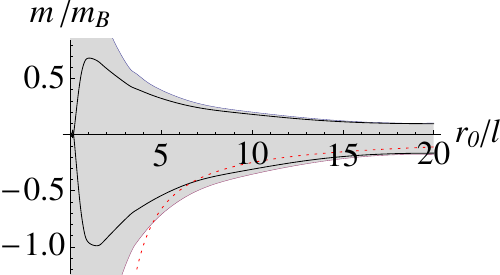}\hskip.5cm\includegraphics[scale=.7]{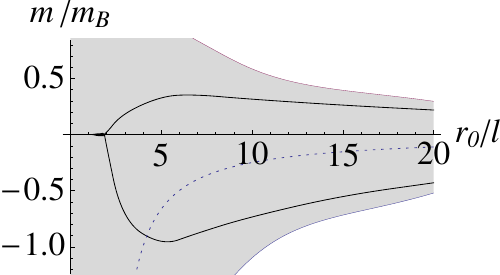}

\centerline{(a)\hskip7cm(b)}
\caption{The phase structure of the effective theory with (a): shifted  and (b): smeared regulated Dirac-delta on the plane $(r_0/\ell,m/m_B)$. The dynamics is stable within the shaded region and the solid lines indicates the solution of the renormalization condition, \eq{rencond}. The dotted line belongs to the linearized theory, fixed by the counterterm \eq{massct}.}\label{phasef}
\end{figure}

The phase structure of the effective theory is shown in Fig. \ref{phasef}. The stability region narrows as the cutoff is removed, $\ell\to0$, since the regulator subjects the trajectory to some deformation within the length scale $\Delta s\sim\ell$, inducing a larger value of the loop integral in the effective equation of motion, \eq{fulleom}, and requiring smaller coefficient, $r_{0B}$. The non-monotonous behavior of the renormalized trajectory indicates the presence of an IR and an UV scaling regime, separated by the intrinsic length scale, $r_0$. There are two solutions of the renormalization condition, one with $m_B>0$ and another with $m_B<0$. The latter is qualitatively consistent with the linearized equation of motion and displays ``Zitterbewegung'', fast oscillations. The non-linear terms of the equation of motion play an important role at any value of the cutoff since the linearized theory is unstable despite having bare parameters within the stability region of the full equation. There is no numerical evidence of a Landau-pole, an obstruction of the limit $\ell\to0$.

\begin{figure}
\includegraphics[scale=.7]{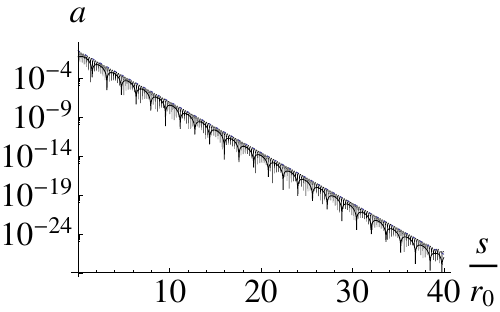}\hskip.5cm\includegraphics[scale=.7]{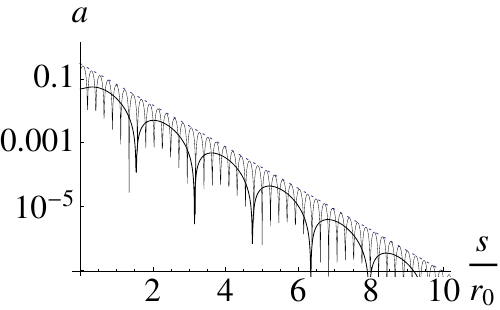}

\centerline{(a)\hskip7cm(b)}
\caption{(a): The spatial acceleration, $|\ddot{\ve{x}}|r_0$, plotted against the proper time, $s/r_0$ for the smeared Dirac-delta regularization at $r_0/\ell=3$ (fat line) and $r_0/\ell=15$ (thin line) with $m_B<0$, plotted together with the prediction of the renormalization condition (dotted line). (b): The zoom is into a more restricted scale region.}\label{scalingsf}
\end{figure}

The quality of satisfying the renormalization conditions is shown in Fig. \ref{scalingsf} (a). While the oscillatory motion of $\ddot{\ve{x}}$ generates strongly localized minima in $|\ddot{\ve{x}}(s)|$ in a periodic manner the envelope follows the prediction of the Abraham-Lorentz force with a remarkable precision despite the non-linearity of the equation of motion. The zoom into Fig. \ref{scalingsf} (a), shown in Fig. \ref{scalingsf} (b), supports the expectation that the length of an oscillation scales with the cutoff. Similar behavior can be found for $m_B>0$ where the monotonous trajectory shows a single exponential relaxation. 

\section{QED}\label{saddlepds}
The physics around the classical electron radius is deeply within the quantum domain and special attention must be paid to justify the use of classical equations of motion in this regime. This issue is taken up here, followed by the discussion of the extention of the saddle point result to a systematic approximation scheme, the inclusion of the radiative corrections.

\subsection{Scale hierarchy and subclassical physics}\label{scaless}
Weakly coupled quantum field theories have an intrinsic hierarchy of scales, assured by the smallness of the dimensionless strength of interaction. In the case of QED with electrons four of the scales  $r_n=\alpha^nr_0$, $\alpha=e^2/\hbar c$, have already been identified, the Bohr radius, $r_{-2}=a_0=\hbar^2/me^2$, the Compton wavelength, $r_{-1}=\lambda_C=\alpha a_0=\hbar/mc$, the classical electron radius, $r_0=\alpha^2a_0=e^2/mc^2$ and finally the Lamb shift scale, $r_1=\alpha^3a_0=e^4/\hbar mc^2$.  The scale dependence of the dynamics is driven by different elementary processes at different scale regimes. This can easily be seen at the first three scales by identifying the fundamental constant which is missing from the expression of the scales. In fact, the driving force comes from the non-relativistic quantum mechanics at the Bohr radius (absence of $c$), from pair creation at the Compton wavelength, independently of the specific nature of the underlying interactions (absence of $e$) and from classical electrodynamics at the classical electron radius (absence of $\hbar$). The physics of the Lamb shift at $r_1$ is driven by involved vacuum polarization effects. The scales with $n\ge2$ and $n\le-3$ are covered by the Electro-Weak theory and the collective phenomena in many-body systems or chemistry, respectively and it is not easy to identify them.

The classical electron radius is well below the quantum-classical transition scale and the dominance of the scaling laws by classical physics is to be taken with a grain of salt. The best is to look at this issue within the context of the saddle point expansion, based on a dimensionless small parameter, $\lambda$, appearing in QED when the rescaling, $\hbar\to\lambda\hbar$ which shifts the quantum-classical transition scales, is performed. This induces the change $\alpha\to\alpha/\lambda$, showing that the saddle point and the usual weak coupling expansion represent two opposite extrema. In fact, we have $r_n\to\lambda^{-2n}r_n$, the gradual turning on of the quantum fluctuations by moving $\lambda$ from 0 to 1 reshuffles the scale hierarchy: The imaginary world with weak quantum fluctuations and large fine structure constant, $a_0<\lambda_C<r_0<r_n$, $n>0$ and our world, $r_n<r_0<\lambda_C<a_0$, are separated by a strongly coupled regime without scale separation, $r_n\sim r_0\sim\lambda_C\sim a_0$ at $\lambda\sim1/137$. Since the separation of scales is an important ingredient of asymptotic expansions one expects complications in extrapolating from the weak to the physical quantum fluctuations. However it is fair to say that the strong field of a point charge creates an $\ord{\hbar^0}$ saddle point contribution around $r_0$, embedded deeply within the quantum domain. 

But the classical limit of a quantum system is more than the recovery of some classical equations of motion. It is instructive in this respect to consider an extension of this problem, the expectation value of local operators in quantum field theory. These expectation values define space-time dependent functions which satisfy integro-differential equation of motion and thus can superficially be viewed as classical fields of a classical effective dynamics as a reminiscent of Ehrenfest's theorem. However the local field variable is classical only if its reduced density matrix is strongly peaked on the diagonal matrix elements. The off-diagonal values characterize the importance of the linear superposition in the averages and must be negligible in the classical domain. Actually it is better to call the expectation values of local operators subclassical fields \cite{sub}, the difference between them and the classical fields being the worse space-time resolution, i.e., lower UV cutoff, and the strong decoherence in the latter case. 

Returning to the the physics of the classical electron radius: The radiation reaction problem at the classical electron radius scale is dominated by the saddle point of the classical radiation of a single point charge, considered to follow a given trajectory. However the interactions between fully dynamical charges in this regime remain obviously dominated by quantum effects.

\subsection{Effective charge dynamics}\label{effactpds}
The motion of a charged particle can be reconstructed from the expectation value of the charge density, thus we seek the relativistic effective theory for the electric current. The aim is limited to show that an UV finite effective action can in principle be constructed by including the perturbative corrections rather than carrying out such a calculation. But there is another reason to follow the argument carefully, it is to underline that the free part of the effective action, controlling the dynamics of the current of the free Dirac see  is a highly involved functional, to be replaced by a phenomenological expression in dealing with the radiation reaction problem. 

The effective action is constructed from the generator functional for the connected Green's functions of the electric current \cite{sub}, 
\be\label{genfop}
e^{\ih W[\hat a]}=\Tr\left[U[a^+,\bar\eta^+,\eta^+]|0\ra\la0|U^\dagger[a^-,\bar\eta^-,\eta^-]\right]
\ee
where the source $a_\mu$, coupled linearly to $A_\mu$ generates the Green's functions and $\bar\eta$, and $\eta$ are coupled linearly to $\psi$ and $\psib$, respectively to produce the coherent initial charged states. In deriving the path integral representation of this functional by applying the usual slicing of the time evolution independently to the operators $U$ and $U^\dagger$ one encounters the integration over pairs of trajectories, $\hat A=(A^+,A^-)$, $\bar\psi=(\psi^+,\psi^-)$ and $\hat{\bar\psi}=(\bar\psi^+,\bar\psi^-)$, coupled to the pairs of sources, $\hat a=(a^+,a^-)$, $\hat\eta=(\eta^+,\eta^-)$ and $\hat{\bar\eta}=(\bar\eta^+,\bar\eta^-)$,
\bea\label{currgenf}
e^{\ih W[\hat a]}&=&\int D[\hat\psi]D[\hat{\bar\psi}]D[\hat A]\exp\ih\biggl[S_M[\hat A]+S_D[\hat{\psib},\hat\psi]\nn
&&+S_i[\hat{\psib}_B,\hat\psi_B,\hat A_B+\hat a]+\int dx[\bar{\hat\eta}(x)\hat\psi(x)+\hat\psib(x)\hat\eta(x)]\biggr].
\eea
The first term,
\be\label{maxact}
S_M[\hA]=\frac1{2c}\int dxdy\hA(x)\hD^{-1}_{Cl}(x,y)\hA(y),
\ee
is the Maxwell action with a relativistic gauge fixing term,
\be\label{dirwa}
\hD_{Cl}^{-1\mu\nu}=-4\pi\left[T^{\mu\nu}+\xi\left(1-h\frac{\Box}{\Lambda^2}\right)L^{\mu\nu}\right]\hD_0^{-1}
\ee
where the notations $L^{\mu\nu}=\partial^\mu\partial^\nu/\Box$ and $T^{\mu\nu}=g^{\mu\nu}-L^{\mu\nu}$ are used with the CTP propagator of a scalar particle of mass $m$, 
\bea\label{ctpprop}
\hD_m(p)&=&\begin{pmatrix}D^n(p)+iD^i(p)&-D^f(p)+iD^i(p)\cr D^f(p)+iD^i(p)&-D^n(p)+iD^i(p)\end{pmatrix}\nn
&=&\begin{pmatrix}\frac1{p^2-m^2+i\epsilon}&-2\pi i\delta(p^2-m^2)\Theta(-p^0)\cr
-2\pi i\delta(p^2-m^2)\Theta(p^0)&-\frac1{p^2-m^2-i\epsilon}\end{pmatrix}.
\eea
Only the longitudinal modes of the gauge field is regularized here, $\Lambda$ and $h>0$ being the cutoff and a regulator parameter, respectively, the transverse sector is regulated later. The CTP propagator contains the Feynman propagator, $D^{++}$, and the retarded and advanced Green's functions, $D^{\stackrel{r}{a}}=D^n\pm D^f$. The free Dirac action,
\be
S_D[\hat{\psib},\hat\psi]=\frac1c\int dxdy\hat{\psib}(x)\hG^{-1}(x,y)\hat\psi(y),
\ee
involves the inverse of the electron propagator $\hG_m(p)=(p\br+m)\hD_m(p)$. The interaction is described by the action
\be
S_i[\hat{\psib},\hat\psi,\hat A]=-\frac{e}c\sum_\sigma\sigma\int dx\psib^\sigma(x)A\br^\sigma(x)\psi^\sigma(x).
\ee
The CTP symmetry of the action, 
\be\label{ctpsym}
S[\phi^+,\phi^-]=-S^*[\phi^-,\phi^+],
\ee
$\phi^\pm$ denoting the CTP doublet pair of a generic field variable, follows from the unitarity of the time evolution.

\subsection{Regularization}\label{regs}
The point splitting has already been used for gauge theories \cite{psplittinggt}, and the particular regularization, implemented here is the replacement of the local fields with smeared ones, $\hat A_B=\hat\kappa\hat\sigma\hat A$, $\hat\psi_B=\hat\chi[\hat A]\hat\sigma\hat\psi$ and $\hat{\psib}_B=\hat{\psib}\hat\sigma\hat{\bar\chi}^{-1}[\hat A]$ with $\bar\chi=\gamma^0\chi^\dagger\gamma^0$, in the interaction where $\hat\sigma=\mr{Diag}(1,-1)$ denotes the simplectic metric tensor of the CTP scheme \cite{psplithd}. The smeared photon field contains the transverse component only, $\hat\kappa^{\mu\nu}=T^{\mu\nu}\hat\kappa_T$, the longitudinal components being suppressed in the interactions. The action is kept invariant under gauge transformations, $A\to A+\partial\alpha$ and $\psi\to e^{-ie\alpha}\psi$, by applying the replacement, $\partial_\mu\to\partial_\mu+ieLA_\mu$, within the smearing function, $\hat\chi$. The physical, gauge invariant components of the electromagnetic field do not appear in the smearing function of the charged field and the potentially dangerous regulator vertices of the higher order derivative scheme \cite{bakeyev} are avoided. The longitudinal sector is regulated by the higher order derivative term of the kernel of the quadratic action for the gauge field in \eq{dirwa}, containing the parameter $h$. 

It is advantageous to perform the change of integral variable, $\hat\psi\to\hat\psi_B$, $\hat{\bar\psi}\to\hat{\bar\psi}_B$, $\hat A\to\hat A_B$, in the generator functional, \eq{currgenf}, which amounts to the replacement $\hD_{Cl}\to\hD_{ClB}=\hat\kappa\hD_{Cl}\hat\kappa$ and $\hG\to\hG_B=\chi\hG_B\bar{\hat\chi}$. The action, expressed in terms of the smeared, bare fields, displays local interaction and modified free dispersion relations \cite{stab}. The sources $\hat\eta,\bar{\hat\eta}$ remain unchanged and they generate the Green's functions of the non-physical bare, smeared fields. Note that the dynamics of these fields is governed by higher order derivative action and as a result is not compatible with the probabilistic interpretation. This is not a problem in the calculation of section \ref{cleffths} where a single point charge follows a given world line. If the charge dynamics is sought in its full complexity then the rescaling of the sources, $\bar{\hat\eta}\to\bar{\hat\eta}\chi^{-1}$ and $\hat\eta\to\bar{\hat\chi}^{-1}\hat\eta$ should be performed to generate the Green's functions of the original field. The latter dynamics remains physically interpretable because the multiplicative factors, $\chi^{-1}$ and $\bar{\hat\chi}^{-1}$, appearing at the external legs remove the non-physical poles of the higher derivative theory.

The regularization of the retarded Green's function, used in section \ref{cleffths}, can be extended to the whole CTP Green's function. Owing to the positivity of the energy of the excitations the support of the spectral function, $iD^{-+}$, is over positive energy, $iD^i(q)=\sign(q^0)D^f(q)$, resulting the Feynman propagator $D^{++}(q^0,\ve{q})=D^r(|q^0|,\ve{q})$. For instance, the retarded Green's function, defined by eq. \eq{shdd} is 
\be
D_B^r(q)=-\frac1{|\ve{q}|}\int_0^\infty dr\frac{r}{r_\ell}e^{ir_\ell(q^0+i\epsilon)}\sin|\ve{q}|r,
\ee
with $r_\ell=\sqrt{\ell^2+r^2}$. Thus the free Green's functions is $\ord{|p^2|^{-\frac32}}$ in the UV regime and renders the Feynman graphs with internal photon lines finite. The CTP matrix, $\hat\kappa$, is assumed to have the block structure of the propagator, \eq{ctpprop}, with $\kappa^{++}(q)=\sqrt{D_B^{++}(q)/D^{++}(q)}$ and $\kappa^r(q)=\sqrt{D_B^r(q)/D^r(q)}$ and $\kappa^i=0$ is chosen to preserve the hermicity of the gauge field. The choice $\hat\eta=\sqrt{-\Lambda\hG_\Lambda}$, with $\Lambda=1/\ell$ and the replacement $\partial_\mu\to\partial_\mu+ieLA_\mu$, preserves causality \cite{stab}.

\subsection{Effective world line action}\label{wlacts}
The perturbation series is generated by the equation
\be
e^{\ih W[\hat a]}=e^{i\frac{\hbar e^2}{2c}\int dxdy\fd{}{\hat a(x)}(x)\hD_{BCl}\fd{}{\hat a(y)}}e^{\ih W_0[\hat a]},
\ee
where the free generator functional is given by
\be
e^{\ih W_0[\hat a]}=\int D[\hat{\bar\psi}]D[\hat\psi]e^{\frac{i}{\hbar c}\int dx\hat{\bar\psi}[\hG_B^{-1}+\hat\sigma\hat a\bre]\hat\psi+\ih\hat{\bar\eta}\hat\psi+\ih\hat{\bar\psi}\hat\eta}.
\ee
The Gaussian integration leads to
\be
W_0[\hat a]=-\hat{\bar\eta}\frac1{c(\hG_B^{-1}-\hat\sigma\hat a\br)}\eta-i\hbar\Tr\ln[\hG_B^{-1}-\hat\sigma\hat a\br],
\ee
the sum of a tree-level and a quantum fluctuation contribution. We shall use the powers of $\hbar$ to trace the weight of the quantum fluctuations hence the Compton wavelength of the electron $mc/\hbar$, the mass term in $\hG$ must be considered as a fixed, $\hbar$-independent number. The effective action for the electric current is given by the Legendre transformation,
\be
\Gamma[\hat j_B]=W[\hat a]-\hat a\hat j_B,~~~\hat j_B=\fd{W[\hat a]}{\hat a}
\ee
and the Euler-Lagrange equation,
\be
\fd{\Gamma[\hat j_B]}{\hat j_B}=\hat a
\ee
is satisfied by $j_B$.

A Gaussian integral can be reproduced by solving the saddle point equation for integration the variable, 
\be
\int_{-\infty}^\infty dAe^{i\frac{D}2A^2+iJA}=e^{-i\frac{J^2}D}\int_{-\infty}^\infty dAe^{i\frac{D}2A^2},
\ee
the linear equation of motion can be used as an exact operator equation and the $\ord{\hbar^0}$ contribution to the generator functional of the connected Green's functions,
\be
W[\hat a]=W_0[\hat a]-\frac{e^2}{2c}\int dxdy\hat j_B(x)\hD^{(T)}_{ClB}(x-y)\hat j_B(y)+\ord\hbar,
\ee
is a saddle point contribution and the quantum corrections come from the polarization of the free Dirac see. This expression contains only the transverse part of photon Green's function, $\hD^{(T)}_{ClB}$, due to the current conservation. The effective action for the current is therefore
\be\label{stfwfac}
\Gamma[\hat j_B]=\Gamma_0[\hat j_B]-\frac{e^2}{2c}\int dxdy\hat j_B(x)\hD^{(T)}_{ClB}(x-y)\hat j_B(y)+\ord\hbar,
\ee
where 
\be
\Gamma_0[\hat j_B]=W_0[\hat a]-\hat a\hat j_B
\ee
stands for the effective action in the free Dirac-see. 

The free effective action is highly involved, displaying a non-local, non-polynomial structure without a small parameter to organize an expansion \cite{ed}. Rather than seeking an approximate solution we assume that it approaches the form,
\be
\Gamma_0[\hat j_B]=-m_Bc\int ds(\sqrt{\dot x^{+2}}-\sqrt{\dot x^{-2}})
\ee
in the point-like limit,
\be
j^{\sigma\mu}_B(x)=\fd{W_0[\hat a]}{\hat a(x)}\to\int ds\delta(x-x^\sigma(s))\dot x^{\sigma\mu}(s),
\ee
for a single charge without pair creation, $\dot x^0>0$. Note that the original, un-smeared current approaches the same limit as $\Lambda\to\infty$. The action \eq{stfwfac} is the CTP extension of the action-at-a-distance theory \cite{sctefo,wheeler}, including the retarded radiation field \cite{point}. The tree-level effective action,
\bea\label{inflfe}
\Gamma[\hx]&=&-m_Bc\int ds(\sqrt{\dot x^{+2}}-\sqrt{\dot x^{-2}})\nn
&&-\frac{e^2}{2c}\sum_{\sigma\sigma'}\sigma\sigma'\int dsds'\dot x^{\sigma,\mu}(s)D_{ClB}^{(T)\sigma\sigma'}(x^\sigma(s)-x^{\sigma'}(s'))\dot x^{\sigma'}_{\mu'}(s'),
\eea
contains the one-loop electron self interaction. This latter is of $\ord{\hbar^0}$ because the electron line of the corresponding Feynman graph describes a coherent state and is $\ord{\hbar^0}$. The solution of the  equation of motion of the action \eq{inflfe} is obtained in section \ref{cleffths} by solving \eq{meom} using the induced gauge field \eq{indgf}.

The form $\Gamma[\hx]=\Gamma_1[x^+]-\Gamma_1[x^-]+\Gamma_2[\hx]$, of the effective action with real $\Gamma_1$ and  $\Gamma_2[x^+,x^-]=-\Gamma^*_2[x^-,x^+]$ is consistent with the symmetry \eq{ctpsym}. The unitarity of the time evolution in the full QED implies $W[a,a]=0$ in eq. \eq{genfop} and consequently $\Gamma[x,x]=\Gamma_2[x,x]=0$. The expectation value $\la x\ra$ is identical when calculated by the help of $U$ or $U^\dagger$ in using the generator functional and the solution of the equations of motion produces $x^+=x^-$. The equation of motion for $x^+$, if $x^+=x^-$,
\be
0=\fd{\Gamma_1[x]}{x}+\fd{\Gamma_2[x,x']}{x}_{|x'=x},
\ee
is identical of eq. \eq{fulleom} and shows the dissipative nature of the radiation reaction force \cite{dec}.

\section{Conclusions}\label{concls}
The scaling laws of an electron around the classical electron radius, $r_0=2.8fm$, deeply within the quantum regime, are governed by the classical equation of motion and the dominant force arises from the interaction of the electron with its own field. The divergent near field of a point charge makes the standard regularization procedure necessary. The effective dynamics of a point charge is derived in this work by the help of a point splitting regularization which smears the electromagnetic field over the invariant invariant distance $ds^2=\ell^2$. The linearized equation of motion describes a stable, causal dynamics for $\ell\gg r_0$ and cutoff-dependent instability arises if $\ell\ll r_0$. However the non-linear terms of the equation of motion owing their existence to the cutoff stabilize the dynamics. Two different renormalized trajectories are found and one of them fits qualitatively to the dynamics of the unstable, linearized equation. The removal of the cutoff seems to be numerically possible, there is no evidence of a Landau pole within the tree-level renormalization. The radiation reaction can be fit into QED as a tree-level saddle point effect.

The historical name of the scale $r_0$ expresses expectations that the classical electrodynamics of a point charge is ill-defined at shorter distances \cite{landau}. One should at this point distinguish two different inquiries. Are we looking into the physics of an imaginary world with classical physics only, $\hbar=0$, or into a physical phenomenon of our world where $\hbar\ne0$? The latter scenarios is followed here by adopting the point of view that classical physics is supposed to be derived from the quantum level and classical electrodynamics should join smoothly to QED at the quantum-classical crossover. Regarding the radiation reaction problem from this point of view one encounters two remarkable features of the Abraham-Lorentz force which are prone to lead to misunderstanding, namely its $\hbar$- and  cutoff-independence. 

While the radiation reaction can be identified in classical electrodynamics and is therefore a purely classical phenomenon however there are three considerations indicating that it is not a typical classical physics problem. First, the radiation reaction force originates from a scale region which is deeply quantum and the correspondence principle, a guiding rule of our intuition, is strongly violated. Second, the tree-level effective equation of motion applies to the expectation value of the world line only, leaving a necessary condition of the classical limit, the decoherence, an open issue. The decoherence, being an IR effect \cite{dec}, is not generated at the scale $r_0$ and the charge maintains its coherent quantum state at this scale, in other words, the Abraham-Lorentz force is a subclassical effect \cite{sub}. The third point concerns the origin and the features of the radiation reaction which bear the fingerprint of quantum field theory, namely being generated by a loop-integral. This integral is divergent and needs a regulator, implying the techniques and concepts of the renormalization group, developed in quantum field theory. 

The radiation reaction force of a point particle is obviously an UV cutoff-effect, the velocity of a massive particle is bounded by the speed of light hence the world line of a point particle can not cut through the light cones of its own radiation field. Among the several cutoff-dependent terms in the effective equation of motion the Abraham-Lorentz force is distinguished by being cutoff-independent. It is generated by the cutoff but its strength is independent of the cutoff scale. This is a well known phenomenon in quantum field theory, has the somehow unfortunate name of anomaly, and reflects the non-uniform convergence of the loop-integrals of the perturbative solutions when the cutoff is removed \cite{dec}. 

The puzzle of the radiation reaction force, the apparent instability of the Abraham-Lorentz force, can be resolved by bearing in mind that the effective classical dynamics contains a one-loop integrals which needs regularization. The introduction of the cutoff makes the parameters of the equation of motion non-physical and forces us to follow the painstaking limit $\ell\to0$ and to construct the renormalized trajectory. The result is a scale-dependent dynamics where it is too naive to expect a simple local differential equation be valid globally, at all scales.

\section*{Acknowledgments}
It is pleasure to thank J\'anos Hajdu for several discussions and Saso Grozdanov for his useful comments in this topics.

\end{document}